\documentclass{vgtc}                          %
\ifpdf%
  \pdfoutput=1\relax                   %
  \usepackage{graphicx}                %
  \DeclareGraphicsExtensions{.pdf,.png,.jpg,.jpeg} %
\else%
  \usepackage{graphicx}                %
  \DeclareGraphicsExtensions{.eps}     %
\fi%

\graphicspath{{figures/}{pictures/}{images/}{./}} %

\usepackage{microtype}                 %
\PassOptionsToPackage{warn}{textcomp}  %
\usepackage{textcomp}                  %
\usepackage{mathptmx}                  %
\usepackage{times}                     %
\usepackage{cite}                      %
\usepackage{tabu}                      %
\usepackage{booktabs}                  %
\usepackage[dvipsnames]{xcolor}
\usepackage{xspace}
\usepackage{comment}
\usepackage{enumitem}

\newcommand{\toolname}{\textsc{MisVis}\xspace}
\newcommand{\misvis}{\textsc{MisVis}\xspace}
\newcommand{\graphview}{Graph View\xspace} 
\newcommand{\summaryview}{Summary View\xspace}
\newcommand{\mainwindow}{Main Window\xspace}
\newcommand{\twitterwindow}{Twitter Window\xspace}
\newcommand{\settingspanel}{Settings Panel\xspace}

\definecolor{controversial-orange}{RGB}{224,130,20}
\definecolor{verified-purple}{HTML}{8a4fbf}
\definecolor{unlabeled-gray}{RGB}{127,127,127}
\newcommand{\controversial}{\textbf{\color{controversial-orange}controversial}\xspace}
\newcommand{\verified}{\textbf{\color{verified-purple}verified}\xspace}
\newcommand{\unlabeled}{\textbf{\color{unlabeled-gray}unlabeled}\xspace}
\newcommand{\www}{World Wide Web\xspace}

\newcommand{\circled}[1]{\raisebox{.5pt}{\textcircled{\raisebox{-.9pt} {#1}}}}
\definecolor{blueVI}{RGB}{30, 136, 229}
\newcommand{\link}[1]{{\href{#1}{\color{blueVI}\textbf{\texttt{#1}}}}}

\definecolor{red}{RGB}{255,0,0}
\newcommand{\revision}[1]{{\color{black}#1}}

\onlineid{0}

\vgtccategory{Research}
\vgtcpapertype{application/design study}

\vgtcinsertpkg

\renewcommand\footnotemark{}

\title{
Explaining Website Reliability by Visualizing Hyperlink Connectivity}
\newcommand{\authorgap}{\hspace{10pt}}
\author{
Seongmin Lee\textsuperscript{\textrm 1} %
  \thanks{\textsuperscript{\textrm 1}Georgia Tech. \{\href{mailto:seongmin@gatech.edu}{seongmin}$\mid$\href{mailto:haekyu@gatech.edu}{haekyu}$\mid$\href{mailto:jayw@gatech.edu}{jayw}$\mid$\href{mailto:oshaikh@gatech.edu}{oshaikh}$\mid$\href{mailto:polo@gatech.edu}{polo}\}@gatech.edu} \authorgap
Sadia Afroz\textsuperscript{\textrm 2}
\thanks{\textsuperscript{\textrm 2}AVAST Software. 
\{\href{mailto:sadia.afroz@gatech.edu}{sadia.afroz}$\mid$\href{mailto:vibhor.sehgal@gatech.edu}{vibhor.sehgal}$\mid$\href{mailto:ankit.peshin@gatech.edu}{ankit.peshin}\}@avast.com} \authorgap
Haekyu Park\textsuperscript{\textrm 1}
\authorgap 
Zijie J. Wang\textsuperscript{\textrm 1}
\authorgap 
Omar Shaikh\textsuperscript{\textrm 1}
\authorgap \\
Vibhor Sehgal\textsuperscript{\textrm 2}
\authorgap
Ankit Peshin\textsuperscript{\textrm 2}
\authorgap 
Duen Horng (Polo) Chau\textsuperscript{\textrm 1}
}

\teaser{
  \centering
  \includegraphics[width=0.859\linewidth]{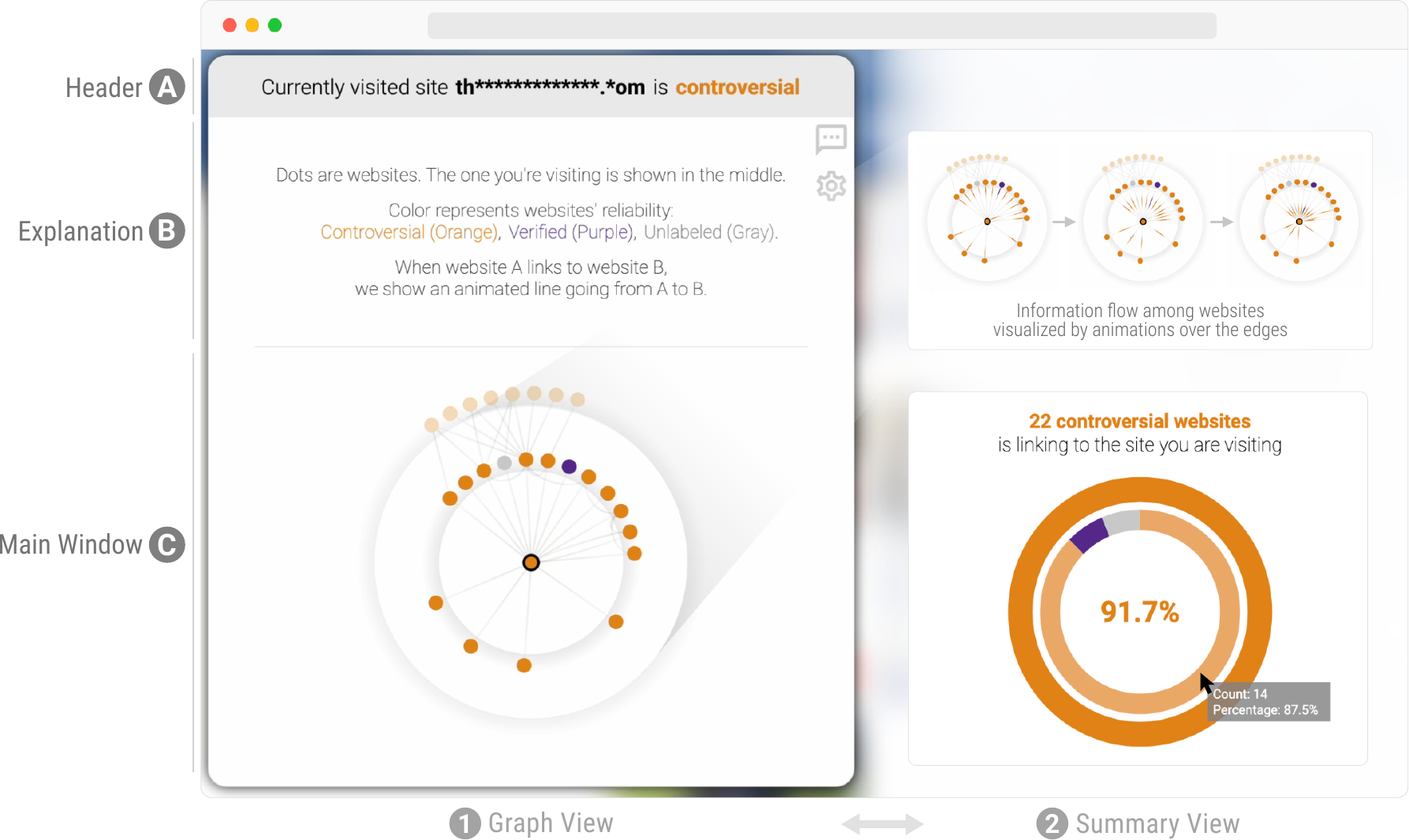} 
  \caption{
  \toolname helps users assess a website's reliability and 
  understand how the site may be involved in spreading false information
  by visualizing its hyperlink connectivity.
  When a user visits a website,
  \toolname is displayed interstitially over the website, blurring its contents and blocking user interactions.
  The \toolname user interface consists of 
  \textbf{(A)} a \textit{Header} message about the reliability of the website being visited by the user,
  \textbf{(B)} brief \textit{Explanation} about how to interpret the \toolname{} visualization, and
  (C) the \textit{\mainwindow} that visualizes the connectivity of the visited website in two coordinated views: \textit{\graphview} and \textit{\summaryview}.
  \textbf{(1)} The \textit{\graphview} shows how the website connects to other websites via hyperlinks. Dots are websites, and edges are hyperlinks; the site being visited is shown in the middle.  
  \toolname{} visualizes the information flow among sites through animation.
  \textbf{(2)} The \textit{\summaryview} shows 
  the site's overall reliability by summarizing the reliability of its connected sites.
  In both views,
  \controversial sites are shown in \textcolor{controversial-orange}{orange},
  \verified in \textcolor{verified-purple}{purple}, and
  \unlabeled in \textcolor{unlabeled-gray}{gray}.
  }
  \label{fig:teaser}
}

\abstract{
As the information on the Internet continues growing exponentially, understanding and assessing the reliability of a website is becoming increasingly important. 
Misinformation has far-ranging repercussions, from sowing mistrust in media to undermining democratic elections.
While some research investigates how to alert people to misinformation on the web,
much less research has been conducted on explaining
how websites engage in spreading false information. 
To fill the research gap,
we present \toolname,
a web-based interactive visualization tool 
that helps users assess a website's reliability by understanding how it engages in spreading false information on the \www.
\toolname visualizes the hyperlink connectivity of the website
and summarizes key characteristics of the Twitter accounts that mention the site.
A large-scale user study with 139 participants demonstrates that \misvis facilitates users to assess and understand false information on the web
and node-link diagrams can be used to communicate with non-experts.
\misvis is available at the public demo link: \link{https://poloclub.github.io/MisVis}.
} %

\CCScatlist{
  \CCScatTwelve{Human-centered computing}{Visu\-al\-iza\-tion}{Visu\-al\-iza\-tion systems and tools}{Visu\-al\-iza\-tion toolkits};
}

\begin{document}

\maketitle

\section{Introduction} %
\label{sec:100:intro}
As the information on the Internet continues growing exponentially,
understanding and assessing the reliability of a website is becoming
increasingly important. 
Misinformation has far-ranging repercussions, from sowing mistrust in media to undermining democratic
elections~\cite{siddiqui2016nc,tasnim2020impact}. 
For example, in the 2016 U.S. election, the top 20 fake news stories on Facebook \textit{had higher engagement than the top 20 true ones} \cite{fakenews2016}. 
Some research investigates how 
to alert people to fake news on the web~\cite{ennals2010highlighting,lanius2021use,kaiser2021adapting}.
Yet, warning about fake news alone does not necessarily help people learn about how to assess information reliability~\cite{pluviano2017misinformation,sharevski2022misinformation}.
Little research has been conducted on explaining how  websites engage in spreading the false information~\cite{patricia2019link,hanley2022no}.

To fill this research gap,
we present
\toolname, a web-based interactive visualization tool that helps users assess a website's reliability 
by understanding how it engages in spreading false information on the web and social media. 
\revision{
Built on top of our early prototype~\cite{lee2022misvis},
we design and develop \toolname following the design study methodology by Sedlmair et al.~\cite{sedlmair2012design}.
} %
\toolname{}'s design is inspired by the recent findings from Sehgal et al.~\cite{sehgal2021mutual}
that the reliability of a website is closely related to
how it is connected to other websites through hyperlinks
and
how it is shared on social media. 
We define \textit{false information} to encompass all types of fake or inaccurate information on the web, such as misinformation, disinformation and conspiracy~\cite{wu2019misinformation}.
Our \toolname{} research contributes:

\begin{itemize}[topsep=2pt, itemsep=0mm, parsep=3pt, leftmargin=10pt]
    \item 
    \textbf{Abstracting the Problem of Explaining Website Reliability.}
    In collaboration with AVAST, a large cybersecurity company,
    we characterize the problem of {explaining website reliability to non-expert users} %
    as the task of visualizing and summarizing websites' hyperlink connectivity and Twitter mentions.
    We design and develop the web-based interactive \toolname{} tool (\autoref{fig:teaser})
    to accomplish that task, helping users better understand
    how websites engage in spreading false information.
    \item 
    \revision{\textbf{Evaluation of Visualization Design of \misvis.}}
    A large-scale user study with 139 participants shows that \misvis effectively helps users identify and understand false information on the web.
    For a demo video of \misvis, visit 
    \link{https://youtu.be/BRp3tedaNeg}.
    \item \textbf{Reflection and Design Lessons.}
    Our iterative design process and study results have made discoveries relating to the larger research area of visualization, adding to the body of knowledge that benefits researchers, e.g., our large-scale user study shows 
    that non-experts can easily comprehend node-link diagrams for understanding hyperlink connectivity; and that displaying interstitial~\cite{kaiser2021adapting} visualization could increase its ease of use. 
\end{itemize}

\section{Related Works}
\label{sec:200:related}
\textbf{Reliability assessment of web information.}
To help people 
assess the factualness of web information,
online platforms, such as Snopes~\cite{snopes}, 
FAIR~\cite{fair}, FactCheck.org~\cite{factcheck}, and PolitiFact~\cite{politifact} focus on helping people manually validate information,
while Ciampaglia et al.~\cite{ciampaglia2015computational} introduces a computational fact-checking technique.
To curb  misinformation on  social media,
Facebook~\cite{facebookmisinfo} and Twitter~\cite{twittermisinfo} have been adding warning labels to alert users to false information,
and web-browser extensions have been developed to detect fake news on social media ~\cite{botsentinel,projectfib}.
However, most existing techniques primarily focus on alerting people to false information~\cite{ennals2010highlighting,lanius2021use,kaiser2021adapting,eccles2021three,kirchner2020countering}.
Much less research has been conducted on explaining how websites engage in spreading false information~\cite{patricia2019link,hanley2022no}.

\smallskip 
\noindent 
\textbf{Hyperlink connectivity visualization.} 
While the visualization community has developed techniques for understanding graph structures~\cite{zhang2021pprviz,south2020debatevis,smith2010nodexl,douma2009spicynodes},
cybersecurity researchers primarily focus on studying hyperlink graphs to better understand  website relationships~\cite{hanley2022no,aoki2020graph,sattar2019detecting}.
However, little research has focused on visualizing  hyperlink connectivity to accomplish the goal of explaining website reliability
~\cite{aoki2020graph}.
Furthermore, to the best of our knowledge, there have not been large-scale studies that evaluate whether non-experts can easily comprehend node-link diagrams for understanding hyperlink connectivity
~\cite{south2020debatevis,douma2009spicynodes}.
To fill the above research gaps, we design, develop, and evaluate \misvis to 
help the general public better understand how websites
engages in spreading false information through visualizing website connections as node-link diagrams.

\section{Design Goals}
\label{sec:300:designgoals}
We have been collaborating closely with security and misinformation domain experts at AVAST since August 2021, iteratively designing and developing \misvis;
we identified three design goals (\ref{goal:visualization}-\ref{goal:accessibility}):

\begin{enumerate}[topsep=2pt, itemsep=0mm, parsep=3pt, leftmargin=19pt, label=\textbf{G\arabic*.}, ref=G\arabic*]
    \item \label{goal:visualization}
    \textbf{Easily Understandable Visualization.}
    While graph visualizations has been developed
    to better understand hyperlink connectivity~\cite{turetken2007visualization},
    there has been little research on whether non-experts can easily understand them~\cite{south2020debatevis}.
    \misvis aims to visualize hyperlink connectivity in a way that is easy for the general public to understand.
    \item \label{goal:credibility} 
    \textbf{Credibility Identification.}
    For users to trust \misvis's visualizations about  websites that they may visit (or to turn away from, as they spread false information), 
    it is important for \misvis to maintain high credibility~\cite{hovland1951influence,brandtzaeg2017trust}.
    We design \misvis to be transparent and neutral in terms of website reliability labels.
    \item \label{goal:accessibility}
    \textbf{Easy to Use.}
    To prevent warnings about false information from being ignored
    ~\cite{kaiser2021adapting,sharevski2022misinformation},
    we aim to design \misvis{}
    \revision{ 
    not to require users any extra efforts to use and understand.
    }
\end{enumerate}

\section{System Design}
\label{sec:400:design}
\subsection{Overview}
\label{subsec:410:overview}
We design
\toolname 
to help users assess a website's reliability 
by  visually explaining how the website engages in spreading false information on the web and social media.
Based on our discussion with AVAST cybersecurity experts and building on 
previous work~\cite{sehgal2021mutual},
we characterize the problem of explaining website reliability to
non-expert users as the task of visualizing and summarizing websites' hyperlink connectivity and Twitter mentions. 
\smallskip
\noindent
\textbf{Dataset.}
For  website reliability and connectivity,
we employ the dataset collected by Sehgal et al.~\cite{sehgal2021mutual},
which consists of 1,059 misinformational and 1,059 informational websites
\revision{that are collected and labeled by combining four publicly available datasets,} %
BS Detector~\cite{bsdetector},
Columbia Journalism Review,
FakeNewsNet~\cite{shu2018fakenewsnet}, and
Media Bias Fact Check~\cite{mediabiasfactcheck}.
We reveal these label sources in \misvis for better transparency (\ref{goal:credibility}).
\toolname labels the misinformational websites as \controversial %
to encompass all types of false information, 
such as misinformation, disinformation, and conspiracy %
(\ref{goal:credibility}).
The informational and unlabeled domains are labeled as \verified and \unlabeled, respectively.

For the Twitter user data,
Twitter's Search Tweets API\footnote{https://developer.twitter.com/en/docs/twitter-api/v1/tweets/search/guides/standard-operators}
has been used with the query ``which websites are shared by which Twitter users''.
Then, the Twitter users who have recently
mentioned at least one of the websites in the website reliability dataset described above %
have been added to the Twitter user dataset.
Among these Twitter users, we identified bot accounts using the botometer-python API\footnote{https://github.com/IUNetSci/botometer-python}.

\smallskip
\noindent
\textbf{User Interface.}
\toolname is displayed interstitially when a user visits a website~\cite{kaiser2021adapting} (\ref{goal:accessibility}).
We implement \misvis using the standard HTML/CSS/JavaScript web technology stack
and the D3.js visualization library~\cite{bostock2011d3}.
\misvis is available at  \link{https://poloclub.github.io/MisVis}.

\subsection{Main Window}
\label{subsec:420:mainwindow}
\misvis's \mainwindow (\autoref{fig:teaser}C)
visualizes the connectivity of the visited website in two coordinated views:
\graphview (\autoref{fig:teaser}\circled{1})
and \summaryview  (\autoref{fig:teaser}\circled{2}).
The \graphview shows how a website is connected with other websites by hyperlinks
and how the information would flow through the links.
The \summaryview presents the visited site's overall reliability  
by summarizing the reliability distributions of its connected sites.

\subsubsection{Graph View}
\label{subsubsec:421:graphview}
The \graphview (\autoref{fig:teaser}\circled{1})
explains how a user-visited website engages in spreading false information. Specifically, it shows how a visited website is connected to other websites by hyperlinks
and how false information would flow through the links.

Each website is represented by a circular node,
whose color indicates the site's reliability;
{\textit{\color{controversial-orange}orange}\xspace} for \controversial,
{\textit{\color{verified-purple}purple}\xspace} for \verified, and
{\textit{\color{unlabeled-gray}gray}\xspace} for \unlabeled.
The label \textit{\unlabeled} is assigned to content aggregators (e.g., \textit{google.com})
or sites whose labels are not yet available.
The nodes are arranged along two concentric rings, based on how they are connected to the visited website.
The website being visited is shown in the center.
The visited site's
1-hop (directly linked)
and 2-hop neighbors
are positioned on the inner and outer ring, respectively.
We include neighbors up to 2 hops away,
because a 2-hop neighborhood provides rich connectivity information for understanding the spread of false information~\cite{sehgal2021mutual,hanley2022no} 
without creating overwhelming visual complexity.
Hyperlinks between two sites are represented as edges.
When a user hovers over a node,
\toolname visualizes how information flows to or from that node.
When 
a \textit{website A} links to \textit{website B},
\toolname shows an animated line going from \textit{A} to \textit{B}.
When none of the nodes is hovered,
the hyperlinks between the visited website and its 1-hop neighbors are animated by default to draw the user's attention to the visited site shown in the center
(\autoref{fig:teaser}\circled{1})

To enhance the readability of the node-link diagram
by arranging the nodes to reduce edge crossings (\ref{goal:visualization}),
we lay out the nodes using the force-directed layout~\cite{fruchterman1991graph} via the
d3-force API\footnote{https://github.com/d3/d3-force}.
Also, we use a straight line to connect two nodes that are on different rings (e.g., connecting a 1-hop neighbor and a 2-hop neighbor)
and
a curved line to connect two nodes that are on the same ring 
(e.g., connecting two 1-hop nodes).

\subsubsection{Summary View}
\label{subsubsec:422:summaryview}

\begin{figure}[t]
    \centering
    \includegraphics[width=0.95\columnwidth]{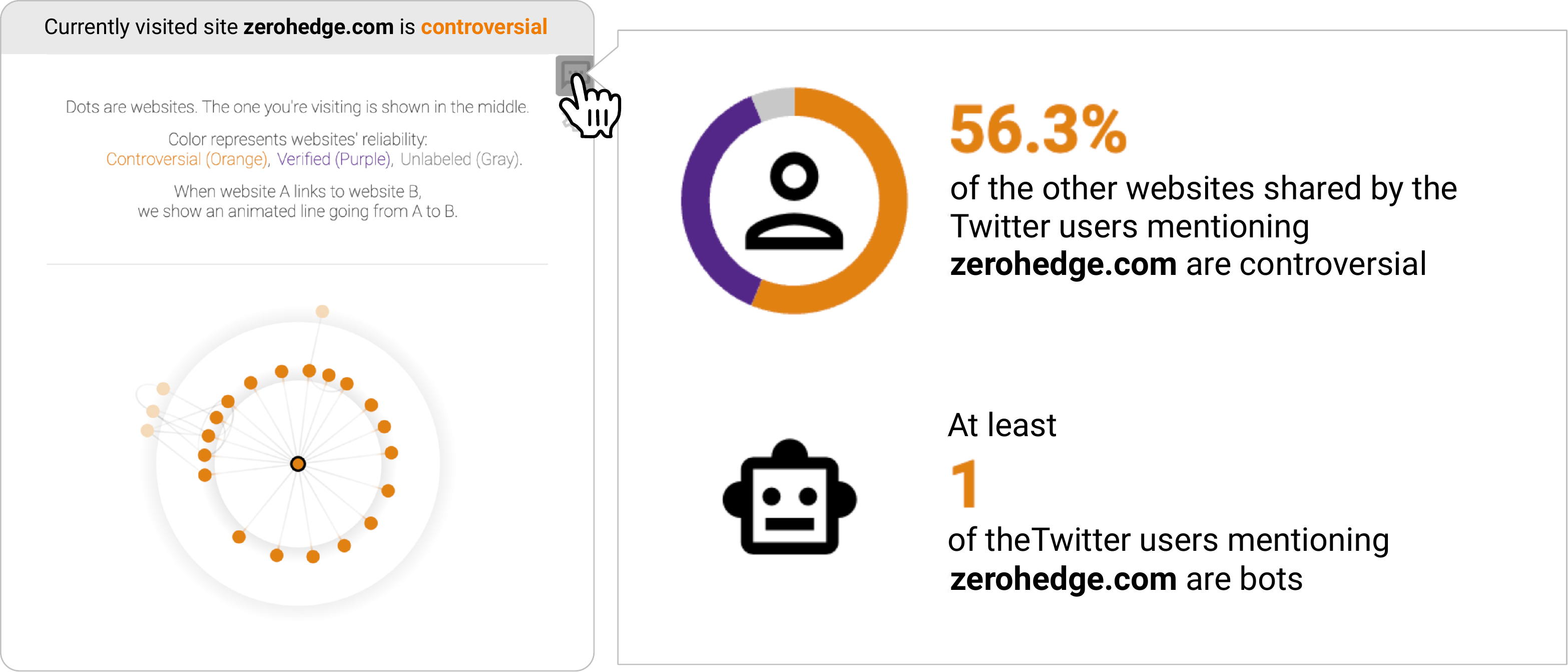}
    \caption{
    \twitterwindow presents two key characteristics of the Twitter accounts that mention the visited website (e.g., zerohedge.com~\cite{hawkins2020twitter,fraser2020google}):
    (1) the reliability distributions of the other sites shared by those Twitter accounts and
    (2) the number of bots among those accounts.
    }
    \label{fig:twitterview}
\end{figure}

The \summaryview (\autoref{fig:teaser}\circled{2})
helps users understand the visited site's  reliability
by providing a quantitative summary of 
its connected sites' reliability.
The \summaryview displays a
\textit{summary statement} (e.g., ``\textbf{\textcolor{controversial-orange}{22 controversial websites}} are linking to the site you are visiting")
and a \textit{doughnut chart} that shows the reliability distributions of the sites linked with the visited site.
As it is common for the sites with false information to link to each other via hyperlinks~\cite{starbird2018ecosystem},
the summary statement raises the user's awareness of the visited site's risk
by highlighting the number of linked \controversial sites.

The doughnut chart summarizes the reliability distributions of its connected sites in two rings;
the inner and outer rings represent the visited site's 1-hop
and 2-hop neighbors, respectively.
In the center of the rings,
we display
the percentage of \controversial sites 
among all the sites in the chart. %
The number and percentage that an arc in the doughnut chart stands for
can be viewed by hovering the arc.

The \summaryview visualizes the neighboring sites' reliability in two modes: \textit{normalized}
and \textit{absolute} mode. %
In the \textit{normalized} mode,
one full ring is for 100\% of the sites in the ring;
for example, 
if 5 out of 10 directly connected sites are \controversial, %
half of the inner ring is colored in \textcolor{controversial-orange}{orange}.
In the \textit{absolute} mode,
each ring is evenly divided into 100 arc segments, 
and each segment corresponds to one website;
for example, 5 \controversial sites are represented by 5 \textcolor{controversial-orange}{orange arc segments}.
We experimented with the number of segments beyond 100 but decided not to use more than 100 segments as they became illegible.
If there are more than 100 sites in a ring, \toolname shows a pop-up message informing that the limit has been reached and reverts to the normalized mode.

\subsection{Twitter Window}
\label{subsec:430:twitterwindow}

As false information
is often shared and propagates on social media~\cite{meel2020fake,naeem2021exploration},
the \twitterwindow (\autoref{fig:twitterview}) 
informs users of how %
the visited website is shared on Twitter.
At the top of the \twitterwindow,
users can see the percentage of the \controversial websites among the sites shared by the Twitter users that have mentioned the visited website.
If the percentage of \controversial sites mentioned by the common Twitter users with the visited website is high,
the visited site is likely to be \controversial as well,
since prolific spreaders of false information
often mention multiple \controversial sites on Twitter.
Below,
the number of bot Twitter accounts that have mentioned the visited website is displayed.
As bot accounts are commonly deployed to spread false information~\cite{metz2020twitter,himelein2021bots},
a high number of bots strongly implies that the site would contain false information.
The \twitterwindow can be shown and hidden
by clicking the social media button
at the top-right corner of the \mainwindow.

\subsection{Settings Panel}
\label{subsec:440:settingspanel}
\begin{figure}
    \centering
    \includegraphics[width=0.98 \columnwidth]{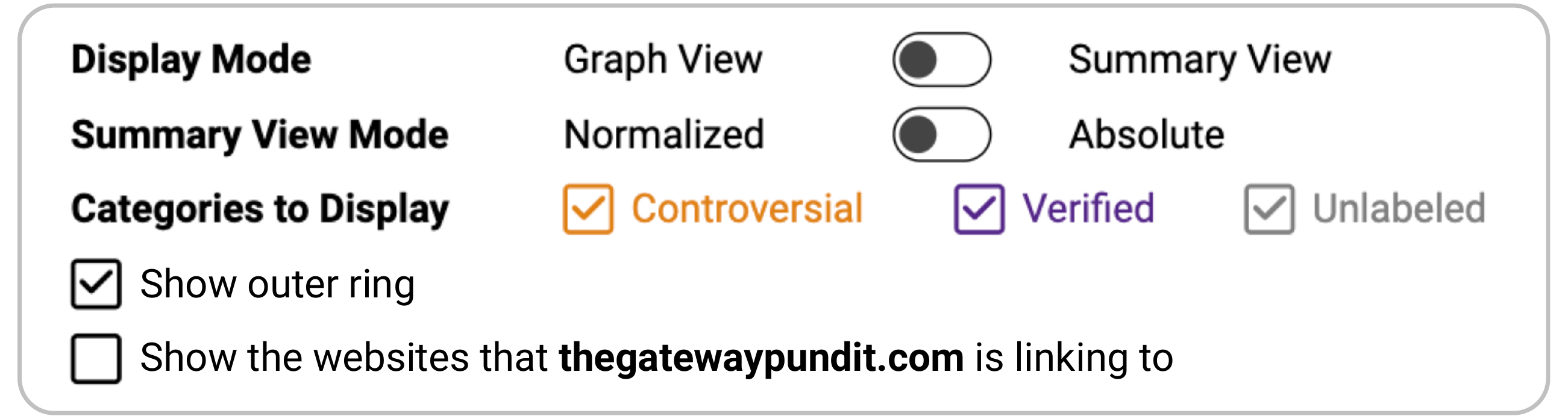}
    \caption{
    The \settingspanel allows users to 
    switch the view of the \mainwindow between the \graphview and the \summaryview;
    switch the \summaryview mode between \textit{normalized} and \textit{absolute};
    select which reliability labels to visualize in the \mainwindow;
    show or hide the outer ring for 2-hop neighbors;
    and show or hide the visualization for the websites that the visited website contains hyperlinks to.
    }
    \label{fig:settingspanel}
\end{figure}

The \settingspanel (\autoref{fig:settingspanel}) is used to %
(1) switch the view of the \mainwindow between the \graphview and the \summaryview (\autoref{subsec:420:mainwindow}),
(2) switch the \summaryview mode
between \textit{normalized} and \textit{absolute}
(\autoref{subsubsec:422:summaryview}),
(3) include or exclude certain reliability labels to or from the visualization in the \mainwindow,
(4) show or hide the outer ring for 2-hop neighbors, and
(5) show or hide the visualization for the
\textit{websites that the visited website contains hyperlinks to},
in addition to the \textit{websites with hyperlinks to the visited website}.
By default,
we display both 1-hop and 2-hop neighbors of the visited website
as 2-hop neighborhood can provide rich information for the connectivity~\cite{sehgal2021mutual,hanley2022no}.
On the other hand, 
the sites that the visited website has hyperlinks to is not shown in the default setting
as 
a \controversial site 
can
deliberately link to a large number of reputable sites
to falsely inflate its credibility and mislead users~\cite{gyongyi2004combating}. %

\section{Design Validation by User Study}
\label{sec:500:study}
To validate the effectiveness of \toolname,
we conducted a large-scale user study.
We recruited 150 U.S.-based participants from Prolific\footnote{https://www.prolific.co/},
an online platform designed for academic research.
The participants' ages range from
``18-24 years old'' to ``65 or older''.
They consume news from a variety of sources,
including news websites, social media, and television.
The study for each participant lasted for around 15 minutes and
we compensated each participant with \$2.50; 
we paid a \$1.00 bonus 
to the participants who provided 
feedback for the open-ended survey questions.

\subsection{Procedure}
\label{sec:510:procedure}
We first asked participants 
to watch a 2-minute tutorial video about how to use \misvis.
After that, to ensure high quality result,
we asked the participants
to answer 3 simple questions about the video.  %
Then, we asked participants to use \toolname in a hypothetical scenario
where they were using a search engine to look up  information to write an essay about the 
\textit{``islands in the East China Sea with territorial dispute''}, %
a topic that is less likely to evoke extreme emotions among most U.S.-based participants.
We provided the participants with the hypothetical search result page showing a list of websites (similar to those returned by search engines).
They were asked to visit the first two websites in sequence.
To prevent exposing the participants to false information and to help them focus on the visualization,
when they clicked on a website,
\toolname was displayed interstitially~\cite{kaiser2021adapting};
the website contents were blurred, user interactions were blocked,  and the website name was masked (\autoref{fig:teaser}).
After visiting each website,
the participants were asked 
to determine if the site was a reliable information source:
\textit{yes}, \textit{maybe}, \textit{no}.

Every participant visited one \controversial site and one \verified site, so that we could
 evaluate how participants interact with \toolname in both cases.
The two websites were presented in random order to guard against order effect (i.e., \controversial$\rightarrow$\verified, or \verified$\rightarrow$\controversial)
We sampled 10 controversial  
and 10 
verified
sites with at least one 1-hop and 2-hop neighbor,
to allow us to evaluate whether the participants could easily use and understand \misvis even with more complex connectivity (\ref{goal:visualization}).

\subsection{Results}
\label{subsec:520:results}
\begin{figure}
    \centering
    \includegraphics[width=1.\columnwidth]{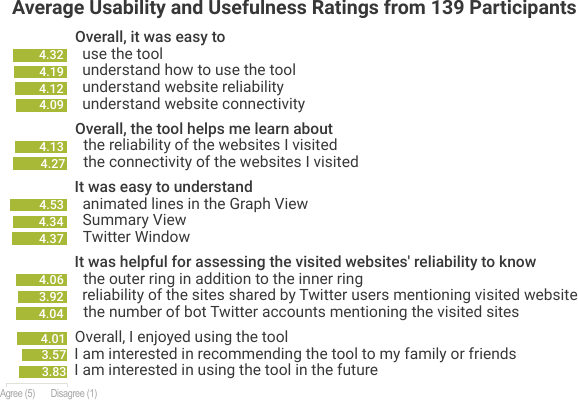}
    \caption{
    Average ratings from 139 participants about 
    \misvis.
    Most participants 
    found \misvis and its features easy to use and helpful.
    }
    \label{fig:userstudy}
\end{figure}

After using \toolname,
the participants were asked 
to answer 5-point Likert-scale questions about \toolname's usability and usefulness for assessing website reliability \cite{jahanbakhsh2021exploring,kaiser2021adapting}.
\autoref{fig:userstudy} summarizes the results from 
the 139 participants
who correctly answered at least two of the three quality-check questions
(we excluded 11 participants who incorrectly answered two or all questions).
Overall, participants rated \misvis favorably. 
They enjoyed using \misvis, and found it helpful and was easy to use and understand.

\smallskip
\noindent
\textbf{Participants could easily use and understand \misvis.} 
The average ratings for all the questions about the overall usability of \misvis and its features
were above 4.
In particular, many participants commented that \misvis was easy to use.
(e.g.,
``\textit{It was laid out nicely and easy to understand.}")

\smallskip
\noindent
\textbf{Node-link diagrams for hyperlink connectivity were easy to understand and helpful (\ref{goal:visualization}).}
The understandability and helpfulness of \toolname's node-link diagram design received high ratings. %
We are excited that non-experts could easily understand such graph based visualization, because this discovery contributes to the body of knowledge of the larger area of visualization, in addition to misinformation research; such a large-scale study has not been performed before.  
Many participants found the animated edges enhanced their understanding.
(e.g., ``\textit{It showed the connections instantly and easily. It was fun to use, and the animation helped easier understanding and was relaxing.}")

\smallskip
\noindent
\textbf{\misvis helped participants assess website reliability and connectivity.}
Many participants gave high ratings to the questions about the helpfulness of \misvis.
After visiting the \controversial websites,
95 out of 139 participants determined the sites to be indeed controversial, 30 were undecided; only 14 thought they were reliable.
For the \verified websites, 
88 determined the sites to be reliable, 40 were undecided.
Our results provided strong empirical evidence for \misvis's effectiveness in helping assess website reliability. 

\smallskip
\noindent
\textbf{Hyperlink connectivity helped participants assess website reliability.} %
Participants commented
that knowing a website's hyperlink connectivity
enhanced their assessment confidence
(e.g.,
``\textit{
Legitimacy of a website's connections bolsters its trust.}",
``\textit{Knowing that a website is connected to reliable websites gives me comfort.}").
This finding validated our problem abstraction of explaining website reliability via visualizing hyperlink connectivity.

\smallskip
\noindent
\textbf{Interstitial visualization enhances \misvis{}'s ease of use (\ref{goal:accessibility}).}
\misvis was displayed interstitially~\cite{kaiser2021adapting},
over the website's blurred contents.
One of the participants said
``\textit{I found this tool useful as I could see the reliability of the website without clicking anything, therefore save me time}",
echoing the benefit of the interstitial design.

\section{Reflection and Design Lessons}
\label{sec:600:reflection}

\smallskip
\noindent
\textbf{Label Wording for Website Reliability.}
Through close discussion with domain experts, we established three criteria to provide accurate and unbiased labels for website reliability (\ref{goal:credibility}):
(1) the labels should be easily understandable for general users,
(2) they should encompass all types of false information, and
(3) they should be neutral, not to offend users. 
For example,
at first, we labeled each website's reliability as  \textit{misinformation}, \textit{reliable}, or \textit{unlabeled}.
However, the domain experts pointed out that 
the term \textit{misinformation} might not cover all types of false information (e.g., disinformation),
and that users could feel offended
if the sites they often visited were labeled as \textit{misinformation}. 
Thus, we adopted the \controversial label, 
which is more neutral than
\textit{misinformation}.
Likewise, we adopted the \verified label to convey that labels had verifiable  sources.

\smallskip
\noindent
\textbf{Revealing Reliability Label Sources.}
We learned that 
revealing the sources for the website reliability labels
significantly increases the credibility of \misvis~\cite{hovland1951influence,pornpitakpan2004persuasiveness} (\ref{goal:credibility}).
The earlier designs of \misvis
did not mention where the reliability labels came from.
On a pilot study, 
multiple participants were concerned about the potential biases in the reliability labels
and wondered how the labels were determined.
We therefore added a statement at the bottom of the \mainwindow to clearly state the sources: 
``Reliability labels are based on credible sources: Columbia Journalism Review, Media Bias Fact Check, FakeNewsNet." (Clicking the sources would bring the users to their websites.)
In our large-scale study, 
as a result, 
far fewer participants were concerned about the credibility of \misvis.

\section{Conclusion and Future Work}
\label{sec:700:conclusion}
We present \toolname,
a web-based interactive visualization tool
to help users assess the reliability of a website and understand how the website is involved in spreading false information on the \www and social media,
by visualizing hyperlink connectivity of the website and summarizing key characteristics of the Twitter accounts that mention the site.
Through a large-scale user study, we validate that \toolname successfully facilitates users to identify and understand false information on the web.
We plan to deploy \toolname as a web browser extension for broader impact and improved usability.

\clearpage
\bibliographystyle{abbrv-doi}

\bibliography{template}

\begin{thebibliography}{10}

\bibitem{botsentinel}
Bot sentinel.
\newblock \url{https://botsentinel.com}.
\newblock Accessed: 2022-04-28.

\bibitem{factcheck}
Factcheck.org.
\newblock \url{www.factcheck.org}.
\newblock Accessed: 2022-04-28.

\bibitem{fair}
Fair.
\newblock \url{fair.org}.
\newblock Accessed: 2022-04-28.

\bibitem{politifact}
Politifact.
\newblock \url{www.politifact.com}.
\newblock Accessed: 2022-04-28.

\bibitem{mediabiasfactcheck}
Questionable sources.
\newblock \url{mediabiasfactcheck.com/fake-news}.
\newblock Accessed: 2022-04-28.

\bibitem{aoki2020graph}
T.~Aoki and A.~Goto.
\newblock Graph visualization of the dark web hyperlink.
\newblock In {\em 2020 Eighth International Symposium on Computing and
  Networking (CANDAR)}, pp. 89--94. IEEE, 2020.

\bibitem{bostock2011d3}
M.~Bostock, V.~Ogievetsky, and J.~Heer.
\newblock D$^3$ data-driven documents.
\newblock {\em IEEE transactions on visualization and computer graphics},
  17(12):2301--2309, 2011.

\bibitem{brandtzaeg2017trust}
P.~B. Brandtzaeg and A.~F{\o}lstad.
\newblock Trust and distrust in online fact-checking services.
\newblock {\em Communications of the ACM}, 60(9):65--71, 2017.

\bibitem{ciampaglia2015computational}
G.~L. Ciampaglia, P.~Shiralkar, L.~M. Rocha, J.~Bollen, F.~Menczer, and
  A.~Flammini.
\newblock Computational fact checking from knowledge networks.
\newblock {\em PloS one}, 10(6):e0128193, 2015.

\bibitem{douma2009spicynodes}
M.~Douma, G.~Ligierko, O.~Ancuta, P.~Gritsai, and S.~Liu.
\newblock Spicynodes: Radial layout authoring for the general public.
\newblock {\em IEEE Transactions on Visualization and Computer Graphics},
  15(6):1089--1096, 2009.

\bibitem{eccles2021three}
D.~A. Eccles and T.~Dingler.
\newblock Three prophylactic interventions to counter fake news on social
  media.
\newblock {\em arXiv preprint arXiv:2105.08929}, 2021.

\bibitem{ennals2010highlighting}
R.~Ennals, B.~Trushkowsky, and J.~M. Agosta.
\newblock Highlighting disputed claims on the web.
\newblock In {\em Proceedings of the 19th international conference on World
  wide web}, pp. 341--350, 2010.

\bibitem{fraser2020google}
A.-M. Fraser.
\newblock Google bans website zerohedge from its ad platform over comments on
  protest articles.
\newblock {\em NBC News}, 2020.

\bibitem{fruchterman1991graph}
T.~M. Fruchterman and E.~M. Reingold.
\newblock Graph drawing by force-directed placement.
\newblock {\em Software: Practice and experience}, 21(11):1129--1164, 1991.

\bibitem{gyongyi2004combating}
Z.~Gyongyi, H.~Garcia-Molina, and J.~Pedersen.
\newblock Combating web spam with trustrank.
\newblock In {\em Proceedings of the 30th international conference on very
  large data bases (VLDB)}, 2004.

\bibitem{hanley2022no}
H.~W.~A. Hanley, D.~Kumar, and Z.~Durumeric.
\newblock No calm in the storm: Investigating qanon website relationships.
\newblock {\em ICWSM}, 2022.

\bibitem{hawkins2020twitter}
D.~Hawkins.
\newblock Twitter bans zero hedge account after it doxxed a chinese researcher
  over coronavirus.
\newblock {\em The Washington Post}, 2020.

\bibitem{himelein2021bots}
M.~Himelein-Wachowiak, S.~Giorgi, A.~Devoto, M.~Rahman, L.~Ungar, H.~A.
  Schwartz, D.~H. Epstein, L.~Leggio, B.~Curtis, et~al.
\newblock Bots and misinformation spread on social media: Implications for
  covid-19.
\newblock {\em Journal of Medical Internet Research}, 23(5):e26933, 2021.

\bibitem{hovland1951influence}
C.~I. Hovland and W.~Weiss.
\newblock The influence of source credibility on communication effectiveness.
\newblock {\em Public opinion quarterly}, 15(4):635--650, 1951.

\bibitem{snopes}
S.~M.~G. Inc.
\newblock Snopes.
\newblock \url{www.snopes.com}.
\newblock Accessed: 2022-04-28.

\bibitem{jahanbakhsh2021exploring}
F.~Jahanbakhsh, A.~X. Zhang, A.~J. Berinsky, G.~Pennycook, D.~G. Rand, and
  D.~R. Karger.
\newblock Exploring lightweight interventions at posting time to reduce the
  sharing of misinformation on social media.
\newblock {\em Proc. ACM Hum.-Comput. Interact.}, 5(CSCW1), apr 2021. doi: {{%
10\hspace{.1pt}\discretionary{.}{%
}{.}\hspace{.4pt}1145\discretionary{/}{%
}{/}3449092}}


\bibitem{kaiser2021adapting}
B.~Kaiser, J.~Wei, E.~Lucherini, K.~Lee, J.~N. Matias, and J.~Mayer.
\newblock Adapting security warnings to counter online disinformation.
\newblock In {\em 30th USENIX Security Symposium (USENIX Security 21)}, pp.
  1163--1180, 2021.

\bibitem{kirchner2020countering}
J.~Kirchner and C.~Reuter.
\newblock Countering fake news: A comparison of possible solutions regarding
  user acceptance and effectiveness.
\newblock {\em Proceedings of the ACM on Human-computer Interaction},
  4(CSCW2):1--27, 2020.

\bibitem{lanius2021use}
C.~Lanius, R.~Weber, and W.~I. MacKenzie.
\newblock Use of bot and content flags to limit the spread of misinformation
  among social networks: a behavior and attitude survey.
\newblock {\em Social Network Analysis and Mining}, 11(1):1--15, 2021.

\bibitem{lee2022misvis}
S.~Lee, S.~Afroz, H.~Park, Z.~J. Wang, O.~Shaikh, V.~Sehgal, A.~Peshin, and
  D.~H. Chau.
\newblock {MisVis}: Explaining web misinformation connections via visual
  summary.
\newblock In {\em Extended Abstracts of the 2022 CHI Conference on Human
  Factors in Computing Systems}. ACM, 2022. doi: {{%
10\hspace{.1pt}\discretionary{.}{%
}{.}\hspace{.4pt}1145\discretionary{/}{%
}{/}3491101\hspace{.1pt}\discretionary{.}{%
}{.}\hspace{.4pt}3519711}}


\bibitem{meel2020fake}
P.~Meel and D.~K. Vishwakarma.
\newblock Fake news, rumor, information pollution in social media and web: A
  contemporary survey of state-of-the-arts, challenges and opportunities.
\newblock {\em Expert Systems with Applications}, 153:112986, 2020. doi: {{%
10\hspace{.1pt}\discretionary{.}{%
}{.}\hspace{.4pt}1016\discretionary{/}{%
}{/}j\hspace{.1pt}\discretionary{.}{%
}{.}\hspace{.4pt}eswa\hspace{.1pt}\discretionary{.}{%
}{.}\hspace{.4pt}2019\hspace{.1pt}\discretionary{.}{%
}{.}\hspace{.4pt}112986}}


\bibitem{facebookmisinfo}
Meta.
\newblock Combating misinformation, 02 2022.

\bibitem{metz2020twitter}
C.~Metz.
\newblock Twitter bots poised to spread disinformation before election.
\newblock {\em The New York Times}, 2020.

\bibitem{projectfib}
M.~C. Nabanita~De, Anant~Goel and Q.~Chen.
\newblock Project fib, 01 2022.

\bibitem{naeem2021exploration}
S.~B. Naeem, R.~Bhatti, and A.~Khan.
\newblock An exploration of how fake news is taking over social media and
  putting public health at risk.
\newblock {\em Health Information \& Libraries Journal}, 38(2):143--149, 2021.

\bibitem{patricia2019link}
V.~Patricia~Aires, F.~G.~Nakamura, and E.~F.~Nakamura.
\newblock A link-based approach to detect media bias in news websites.
\newblock In {\em Companion Proceedings of The 2019 World Wide Web Conference},
  pp. 742--745, 2019.

\bibitem{pluviano2017misinformation}
S.~Pluviano, C.~Watt, and S.~Della~Sala.
\newblock Misinformation lingers in memory: failure of three pro-vaccination
  strategies.
\newblock {\em PloS one}, 12(7):e0181640, 2017.

\bibitem{pornpitakpan2004persuasiveness}
C.~Pornpitakpan.
\newblock The persuasiveness of source credibility: A critical review of five
  decades' evidence.
\newblock {\em Journal of applied social psychology}, 34(2):243--281, 2004.

\bibitem{bsdetector}
M.~Risdal.
\newblock Getting real about fake news, 2016. doi: {{%
10\hspace{.1pt}\discretionary{.}{%
}{.}\hspace{.4pt}34740\discretionary{/}{%
}{/}KAGGLE\discretionary{/}{%
}{/}DSV\discretionary{/}{%
}{/}911}}


\bibitem{twittermisinfo}
Y.~Roth and N.~Pickles.
\newblock Updating our approach to misleading information, 05 2020.

\bibitem{tasnim2020impact}
M.~M.~H. Samia~Tasnim and H.~Mazumder.
\newblock Impact of rumors and misinformation on covid-19 in social media.
\newblock {\em Journal of Preventive Medicine \& Public Health}, 53:171--174,
  2020. doi: {{%
10\hspace{.1pt}\discretionary{.}{%
}{.}\hspace{.4pt}3961\discretionary{/}{%
}{/}jpmph\hspace{.1pt}\discretionary{.}{%
}{.}\hspace{.4pt}20\hspace{.1pt}\discretionary{.}{%
}{.}\hspace{.4pt}094}}


\bibitem{sattar2019detecting}
N.~S. Sattar, S.~Arifuzzaman, M.~F. Zibran, and M.~M. Sakib.
\newblock Detecting web spam in webgraphs with predictive model analysis.
\newblock In {\em 2019 IEEE International Conference on Big Data (Big Data)},
  pp. 4299--4308. IEEE, 2019.

\bibitem{sedlmair2012design}
M.~Sedlmair, M.~Meyer, and T.~Munzner.
\newblock Design study methodology: Reflections from the trenches and the
  stacks.
\newblock {\em IEEE transactions on visualization and computer graphics},
  18(12):2431--2440, 2012.

\bibitem{sehgal2021mutual}
V.~Sehgal, A.~Peshin, S.~Afroz, and H.~Farid.
\newblock Mutual hyperlinking among misinformation peddlers.
\newblock {\em CoRR}, abs/2104.11694, 2021.

\bibitem{sharevski2022misinformation}
F.~Sharevski, R.~Alsaadi, P.~Jachim, and E.~Pieroni.
\newblock Misinformation warnings: Twitter’s soft moderation effects on
  covid-19 vaccine belief echoes.
\newblock {\em Computers \& security}, 114:102577, 2022.

\bibitem{shu2018fakenewsnet}
K.~Shu, D.~Mahudeswaran, S.~Wang, D.~Lee, and H.~Liu.
\newblock Fakenewsnet: A data repository with news content, social context and
  dynamic information for studying fake news on social media.
\newblock {\em arXiv preprint arXiv:1809.01286}, 2018.

\bibitem{siddiqui2016nc}
F.~Siddiqui and S.~Svrluga.
\newblock N.c. man told police he went to d.c. pizzeria with gun to investigate
  conspiracy theory.
\newblock {\em The Washington Post}, 2016.

\bibitem{fakenews2016}
C.~Silverman.
\newblock Viral fake election news outperforms real news on facebook, 2022.

\bibitem{smith2010nodexl}
M.~A. Smith, C.~A., N.~Milic-Frayling, B.~Shneiderman, E.~Mendes~Rodrigues,
  J.~Leskovec, and C.~Dunne.
\newblock Nodexl: a free and open network overview, discovery and exploration
  add-in for excel 2007/2010/2013/2016.
\newblock 2010.

\bibitem{south2020debatevis}
L.~South, M.~Schwab, N.~Beauchamp, L.~Wang, J.~Wihbey, and M.~A. Borkin.
\newblock Debatevis: Visualizing political debates for non-expert users.
\newblock In {\em 2020 IEEE Visualization Conference (VIS)}, pp. 241--245.
  IEEE, 2020.

\bibitem{starbird2018ecosystem}
K.~Starbird, A.~Arif, T.~Wilson, K.~Van~Koevering, K.~Yefimova, and
  D.~Scarnecchia.
\newblock Ecosystem or echo-system? exploring content sharing across
  alternative media domains.
\newblock In {\em Proceedings of the International AAAI Conference on Web and
  Social Media}, vol.~12, 2018.

\bibitem{turetken2007visualization}
O.~Turetken and R.~Sharda.
\newblock Visualization of web spaces: state of the art and future directions.
\newblock {\em ACM SIGMIS Database: the DATABASE for Advances in Information
  Systems}, 38(3):51--81, 2007.

\bibitem{wu2019misinformation}
L.~Wu, F.~Morstatter, K.~M. Carley, and H.~Liu.
\newblock Misinformation in social media: definition, manipulation, and
  detection.
\newblock {\em ACM SIGKDD Explorations Newsletter}, 21(2):80--90, 2019.

\bibitem{zhang2021pprviz}
S.~Zhang, R.~Yang, X.~Xiao, X.~Yan, and B.~Tang.
\newblock Pprviz: Effective and efficient graph visualization based on
  personalized pagerank.
\newblock {\em arXiv preprint arXiv:2112.14944}, 2021.

\end{thebibliography}
\end{document}